\documentclass[12pt,preprint]{emulateapj}
\begin{document}

\parindent=1.0cm

\title{
NGC 3105: A YOUNG CLUSTER IN THE OUTER GALAXY 
\altaffilmark{1,} \altaffilmark{2}}

\author{T. J. Davidge}

\affil{Dominion Astrophysical Observatory,
\\National Research Council of Canada, 5071 West Saanich Road,
\\Victoria, BC Canada V9E 2E7\\tim.davidge@nrc.ca}

\altaffiltext{1}{Based on observations obtained at the Gemini Observatory, which is
operated by the Association of Universities for Research in Astronomy, Inc., under a
cooperative agreement with the NSF on behalf of the Gemini partnership: the National
Science Foundation (United States), the National Research Council (Canada), CONICYT
(Chile), Minist\'{e}rio da Ci\^{e}ncia,
Tecnologia e Inova\c{c}\~{a}o (Brazil) and Ministerio de Ciencia, Tecnolog\'{i}a e
Innovaci\'{o}n Productiva (Argentina).}

\altaffiltext{2}{This research has made use of the NASA/IPAC Infrared Science Archive,
which is operated by the Jet Propulsion Laboratory, California Institute of Technology,
under contract with the National Aeronautics and Space Administration.}

\begin{abstract}

	Images and spectra of the open cluster NGC 3105 have been obtained with 
GMOS on Gemini South. The $(i', g'-i')$ color-magnitude diagram (CMD) constructed 
from these data extends from the brightest cluster members to $g' \sim 23$. 
This is $\sim 4 - 5$ mag fainter than previous CMDs at visible 
wavelengths and samples cluster members with sub-solar masses. 
Assuming a half-solar metallicity, comparisons with isochrones yield 
a distance of $6.6 \pm 0.3$ kpc. An age of at least $32$ Myr is found based on the 
photometric properties of the brightest stars, coupled with the apparent absence of 
pre-main sequence stars in the lower regions of the CMD. The luminosity function 
of stars between 50 and 70 arcsec from the cluster center is consistent with a 
Chabrier (2003, PASP, 115, 763) lognormal mass function. However, 
at radii smaller than 50 arcsec there is a higher specific frequency of the 
most massive main sequence stars than at larger radii. 
Photometry obtained from archival SPITZER images 
reveals that some of the brightest stars near NGC 3105 have excess infrared 
emission, presumably from warm dust envelopes. H$\alpha$ emission is detected in 
a few early-type stars in and around the cluster, building upon previous 
spectroscopic observations that found Be stars near NGC 3105. 
The equivalent width of the NaD lines in the spectra of early type stars is consistent 
with the reddening found from comparisons with isochrones. Stars 
with $i' \sim 18.5$ that fall near the cluster main sequence 
have a spectral-type A5V, and a distance modulus that is consistent with that 
obtained by comparing isochrones with the CMD is found assuming solar 
neighborhood intrinsic brightnesses for these stars.

\end{abstract}


\section{INTRODUCTION}

	Simulations of molecular clouds suggest that stars form in deeply 
embedded filamentary structures that subsequently collapse and merge to form 
star clusters and a diffusely distributed component (e.g. 
Bonnell et al. 2011). Assuming that the majority of stars form in such 
environments then a census of young stars in clusters and in the so-called `field' 
suggests that most clusters must be short-lived (e.g. Lada \& Lada 2003; 
Silva-Villa \& Larsen 2011; Fall \& Chandar 2012). That stars do not form in isolation 
is important not only for interpreting the nature and evolution of star 
clusters but also for understanding the properties of stars in the field. This is 
because the time that stars stay in close physical proximity in their natal 
environments may affect the properties of the objects that ultimately migrate into 
the field, such as the binary fraction (e.g. Parker \& Meyer 2014) and the low mass 
cut-off of the main sequence (MS; e.g. Johnstone et al. 1998; Adams et al. 2004).

	Feedback from massive stars likely plays a key role in cluster evolution. 
Feedback may purge the gas that is the dominant contributor to the total system 
mass early in a cluster's evolution, thereby altering the gravitational potential so 
that stars are no longer bound to the cluster. Feedback is also a potential regulator 
of star formation during embedded phases (e.g. Nakamura \& Li 2014), and may 
trigger star formation in surrounding areas (e.g. Bik et al. 2014; Getman et 
al. 2014). Feedback may also provide re-cycled material to a star-forming 
region if star formation proceeds over timescales that are longer 
than a few Myr (e.g. Palous et al. 2014).

	If feedback from massive stars is a significant influence on cluster evolution 
then the age distribution of embedded clusters should plunge at a time that 
corresponds to the lifetime of massive stars. Studies of clusters with 
minimum masses of at least a few $\times 10^3$ M$_{\odot}$ 
in the nearby spiral galaxy M83 indicate that the oldest embedded clusters 
have ages $\sim 6$ Myr (Hollyhead et al. 2015), which 
is roughly consistent with the lifetime of very massive stars. 
This age also corresponds to the onset of the red supergiant (RSG) evolutionary 
phase in stellar systems, and so marks a time when the intensity of the local 
UV radiation field will drop if very massive stars were initially present. 

	Factors other than feedback almost certainly also affect cluster evolution. 
The kinematic state of a cluster at the time of gas expulsion 
likely plays a significant role in its survivability, in the sense 
that systems that happen to be in a super-virial state at the epoch of gas 
expulsion will have a greater chance of surviving than those in a sub-virial 
state (e.g. Farias et al. 2015). The fate of a cluster may also depend on 
environmental factors such as proximity to massive molecular clouds 
(e.g. Kruijssen et al. 2012), location in a galaxy (e.g. Silva-Villa et al. 2014), 
and the morphological characteristics of the host galaxy (e.g. de Grijs et al. 
2013). 

	Having a large initial mass may not be a guarantee that a cluster will 
survive for more than a few Myr. Rather, the fate of a young cluster may be 
influenced by local conditions within the natal environment. 
Stars in massive star-forming regions may form in a range of 
environments with surface densities that span at least 4 orders of magnitude (e.g. Kuhn 
et al. 2015), and it is the densest sub-structures that might be expected to have the 
highest chances of long-term survival. Not all systems contain such sub-structures. 
For example, the massive star-forming complex W33 lacks compact sub-structures, and 
will likely evolve into a loose association (e.g. Messineo et al. 2015). 

	The star-to-star age dispersion in a cluster is one measure of how long 
star-forming material is retained, and this can be estimated by measuring the 
age difference between the youngest pre-MS (PMS) stars and the oldest MS stars. 
Realistic stellar structure models that span a wide range of masses and evolutionary 
states are required, and the failure to include key physical processes can skew 
age dispersion estimates. For example, models presented by Somers \& Pinsonneault 
(2015) indicate that the failure to account for star spots may cause isochrones to 
detect an erroneous age dispersion in the CMDs of systems that are actually coeval.

	Observations suggest that age dispersions up to a few Myr are common 
in massive young clusters (e.g. Rom\'an-Z\'u\~niga et al. 2015; Zeidler et al. 2015; 
Kuryavtseva et al. 2012). Such a relatively short 
star-forming history is consistent with observations 
that point to a rapid collapse for massive systems (e.g. Banerjee \& Kroupa 
2015) and the timescale over which feedback may start to influence cluster evolution. 
Still, evidence for larger age dispersions has been found in some cases (e.g. 
De Marchi et al. 2011b; Lim et al. 2013). Environment may also play a role in 
determining the duration of star formation. In particular, 
the spread in age in low mass clusters that form in low density environments 
may be significantly larger than in more massive -- and presumably more 
compact -- clusters (Pfalzner et al. 2014). 

	The Galactic disc contains young clusters in a range of environments, and the 
investigation of clusters and their immediate surroundings throughout the Galaxy 
will provide insights into the processes that affect cluster evolution. 
With R$_{GC} = 11.4 \pm 0.6$ kpc (Paunzen et al. 2005), the open cluster NGC 3105 is 
in the outer regions of the Galactic disc. Previous estimates of the age 
and distance of NGC 3105 are summarized in Table 1. The age estimates in Table 1 draw 
on different age diagnostics and wavelengths. The ages measured 
by Sagar et al. (2001) and Paunzen et al. (2005) are based on the photometric 
properties of MS turn-off (MSTO) and post-MS stars at visible wavelengths. 
In contrast, the age estimated by Davidge (2014) is based on the MS cut-off (MSCO) 
and the properties of the cluster PMS sequence in the $(K, J-K)$ 
color-magnitude diagram (CMD).

\begin{deluxetable}{ccl}
\tablecaption{Previous age and distance estimates of NGC 3105}
\startdata
\tableline\tableline
Age & Distance & Reference \\
(Myr) & (kpc) & \\
\tableline
$25 \pm 10$ & $9.5 \pm 1.5$ & Sagar et al. (2001) \\
$20 \pm 5$ & $8.5 \pm 1.0$ & Paunzen et al. (2005) \\
$25 \pm 5$ & $6.8 \pm 0.3$ & Davidge (2014) \\
\tableline
\enddata
\end{deluxetable}

	In the present paper, deep photometric and spectroscopic observations 
of stars in NGC 3105 are discussed. Images obtained with the Gemini 
Multi-Object Spectrograph (GMOS) on Gemini South
are used to construct a CMD that extends from the very brightest cluster 
members to $g' \sim 23$, which is where PMS stars might be expected based on some 
previous age estimates. If the input physics used in stellar 
structure models are correct then age diagnostics that cover 
different wavelengths and a broad range of magnitudes should yield the same age. 
The range of magnitudes covered with the GMOS data 
enables an investigation of cluster age using indicators that span a range 
of masses and evolutionary states.

	In addition to investigating the age, distance, and reddening of NGC 
3105, the GMOS CMD was also used to select targets for spectroscopic 
observation at visible and red wavelengths. Spectra were obtained 
of stars that span a range of brightnesses, and the data are used 
to investigate the line-of-sight interstellar absorption, search for line emission, 
and estimate spectral types. This is the first spectroscopic survey of NGC 3105 
that includes stars that are fainter than the MSTO. Finally, images of NGC 3105 
that were obtained as part of the SPITZER (Werner et al. 2004) GLIMPSE (Benjamin et 
al. 2003) survey are also examined. These data are used to probe the light profile 
of the cluster and investigate the mid-IR properties of bright stars in and around 
NGC 3105. 

	Details of the observations and the steps used to 
remove instrumental signatures from the data are discussed in Section 2. 
The light profile of NGC 3105 obtained from archival SPITZER images is investigated 
in Section 3. The photometric measurements obtained from the GMOS and SPITZER images 
are described in Section 4, and the CMDs and LFs obtained from these data are examined 
in Sections 5 and 6. The stellar spectra are discussed in Section 7, while a 
discussion and summary of the results follows in Section 8. 

\section{OBSERVATIONS \& REDUCTIONS}

\subsection{GMOS Imaging}

	Images and visible/red spectra were recorded 
with GMOS (Hook et al. 2004) on Gemini South 
as part of program GS-2014A-Q-84 (PI: Davidge). The GMOS detector 
when these data were recorded \footnote[3]{The detector in GMOS has 
since been changed.} was a mosaic of three $2048 \times 4068$ EEV CCDs. 
Each 13.5$\mu$m square CCD pixel subtended 0.073 arcsec per side on the sky. 
All images and spectra for this program were recorded with $2 \times 2$ 
pixel binning.

	$g'$ and $i'$ images of NGC 3105 were recorded on the night of UT 
December 31, 2013. The sky conditions were photometric when the data were recorded.
The exposure times and the mean full width half maximum (FWHM) of stellar profiles at 
each exposure time are summarized in Table 2. Short and long exposures were recorded 
to broaden the magnitude range over which photometry could be performed. 
The long exposure $i'$ images were recorded with a five point dither pattern 
that sampled the corners and center of a $10 \times 10$ arcsec square. 

\begin{deluxetable}{clc}
\tablecaption{Summary of GMOS images}
\startdata
\tablewidth{0pt}
\tableline\tableline
Filter & Exposures & FWHM \\
 & & (arcsec) \\
\tableline
$g'$ & $1 \times 1$ sec & 0.7 \\
 & $1 \times 300$ sec & 0.7 \\
$i'$ & $1 \times 1$ sec & 0.5 \\
 & $5 \times 60$ sec & 0.5 \\
\tableline
\enddata
\end{deluxetable}

	In addition to science data, a series of calibration images that are 
required to process the science data were also obtained. Exposures that 
monitor the floating bias level and the static bias pattern of the CCDs (`biases') 
were recorded at the end of the night. Images that measure 
variations in sensitivity due to instrument optics and pixel-to-pixel 
differences in detector sensitivity (`flats'), and interference fringes in $i'$ 
\footnote[4] {Fringing is not significant in the $g'$ images.} (`fringe frames') 
were provided by Gemini as part of the data package for this program. 
The flats were constructed from observations of the twilight sky.

	A standard processing sequence for CCD mosaic imaging at 
visible and red wavelengths was applied to remove instrumental signatures. 
To start, the outputs from the individual CCDs in the science and 
calibration images were multiplied by their amplifier gains. The bias frames were 
subtracted from the gain-corrected images, and the results were then 
divided by the flats. The fringe frame was scaled to match the 
long and short exposure times of the $i'$ science images, and the result was subtracted 
from these data. The long exposure $i'$ images were registered using the centroids 
of stars located across the imaged field as reference points, and the results 
were averaged together to construct a final deep $i'$ image. 
The $g'$ and short-exposure $i'$ images were aligned with the 
combined deep $i'$ image. This required the application of 
modest offsets ($\leq 0.1$ arcsec) to adjust for slight 
shifts in telescope pointing that occur during guiding. The final deep $i'$ 
image is shown in the left hand panel of Figure 1.

\begin{figure*}
\figurenum{1}
\epsscale{1.2}
\plotone{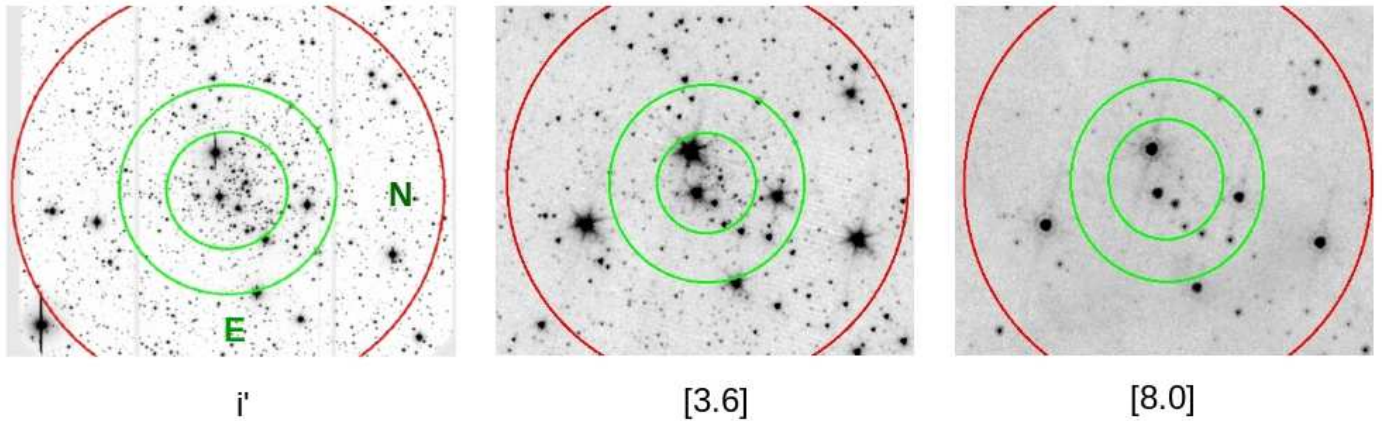}
\caption{Images of NGC 3105 recorded with GMOS and SPITZER. A $5.7 \times 4.6$ arcmin 
field is shown for each filter. The two vertical stripes in the $i'$ image are gaps 
between the CCDs. The green circles mark the outer boundaries of the 
Center and Shoulder regions that are used in the analysis of the photometric and 
spectroscopic data, while the red circle marks the boundary between Fields 1 
and 2.}
\end{figure*}

\subsection{GMOS Spectra}

	Multi-slit spectra of stars in and around NGC 3105 were recorded 
on the night of April 18, 2014. Target selection was based 
on (1) location on the sky close to the center of NGC 3105, 
and (2) location on the $(i', g'-i')$ CMD that 
was constructed from the GMOS images. Targets were restricted to the 
middle CCD of the detector mosaic to maximize common star-to-star wavelength 
coverage. In some cases it was possible to place another source in a slit that 
sampled a high priority target by making minor adjustments to the slit location, 
and these additional sources constitute a modest serendipitous sample. In all, 
spectra were recorded of 35 sources, although the spectrum of one source (\# 22) 
was found to be an artifact of a nearby bright star, and so is not considered further. 
The locations and photometric properties of the stars for which spectra were 
obtained are listed in Table 3.

	With the exception of the three stars used to align 
the focal plane mask (see below), the slits had a length 
of 6 arcsec and a width of 1 arcsec. The light was dispersed with the R400 grating 
(400 l/mm; $\lambda_{blaze} = 7640\AA$), and the spectral resolution measured from 
CuAr emission lines is $\lambda/\Delta \lambda \sim 860$ FWHM. This 
resolution is too low to allow radial velocities that are reliable enough to 
assess cluster membership to be measured. The seeing, estimated from the 
width of stellar profiles in the slits, was $\sim 1$ arcsec FWHM. 

	Three stars were used to align the slit mask on the sky, and 
$2.4 \times 2.4$ arcsec apertures -- rather than slits -- were cut in the mask 
for these sources. These stars are among the brightest in the sample, and 
their light profiles could be examined in short exposure images to check that 
they are centered in their apertures. While spectra were recorded of the 
alignment stars, their spectroscopic line profiles differ slightly from 
those recorded through slits given that their spectral resolution is tied 
to the seeing disk, rather than a 1 arcsec slit. However, 
there is no noticeable difference between the line profiles of 
the alignment stars and other science targets, and so they are considered in 
concert with the slit spectra.

	Four 900 second exposures were recorded, with two having 
a nominal central wavelength of 6800\AA\ and the other two 
a central wavelength of 6850\AA. This offset in central wavelength allows the 
portions of the spectrum that fall on gaps between the CCDs in one wavelength 
setting to be covered with the other setting. 
A GG455 filter was used to suppress higher spectroscopic orders, and 
the combined response of this grating $+$ filter combination 
results in poor performance at wavelengths $< 5000$\AA. 

	Given the broad wavelength coverage, the start and end wavelengths for 
each spectrum depends on the location of the star in the science field, 
and the sources listed in Table 3 have a common wavelength coverage 
between $\sim 5300$ and 8400\AA. This wavelength interval contains atomic 
and molecular transitions that are sensitive to metallicity, temperature, and surface 
gravity. Features that are interstellar in origin are also 
expected in these spectra given that NGC 3105 is viewed through the 
Galactic disk plane, with a line-of-sight extinction A$_V \sim 3$ magnitudes. 
Diffuse Interstellar Bands (DIBs) are clearly evident in the spectra of 
the brightest objects (Section 7.1).

	Flat field frames were recorded through the slit mask at each 
wavelength setting midway through the observing sequence. CuAr arcs were recorded 
through the slit mask at each central wavelength at the end of the night. 
The flat-field and arc light sources are in the Gemini Facility Calibration Unit. 

	The reduction of the spectra proceeded as follows. The individual 
exposures recorded for the science, flat-field, and arc frames were initially processed 
separately, starting with the multiplication by the gain of the output from each CCD. 
A series of bias frames, recorded at the end of the night, were median-combined, and 
the result was subtracted from each exposure.
This was followed by the removal of cosmic rays, which were identified by applying 
a running median filter to highlight features that are sharper than those that result 
from spectroscopic lines and the seeing disk. The different exposures for each 
observation type (science, flat-field, and arc) were aligned to 
correct for the wavelength offsets applied during the 
observing sequence and then averaged together. The null signal in the 
gaps between CCDs was rejected when computing the mean. 

	Individual slitlets were extracted from the combined exposures, and 
each slitlet of a cluster source was divided by the corresponding normalized 
flat-field spectrum. The flat-fielded cluster spectra were 
wavelength calibrated using line identifications made from the arcs. The background sky 
level was measured on a row-by-row basis near the edges of each 
wavelength calibrated slitlet, and the resulting sky spectrum was subtracted 
from each of the columns in that slitlet. The sky level for slits containing two stars 
was measured in the spatial interval where light from the stars was smallest. 
Spectra of individual stars were extracted from the sky-subtracted slitlets by 
combining signal within the FWHM of each stellar light profile. 

	Telluric absorption features are present 
at wavelengths longward of 6600\AA. Spectra of the white dwarf LTT 4364 that were 
recorded for program GS-2014A-Q-12 on May 7, 2014 with the same slit width, 
grating, airmass, and central wavelength as the NGC 3105 observations 
were used to suppress these telluric features. The LTT 4364 spectra were 
processed with the same procedures as the NGC 3105 spectra. 
Division of the NGC 3105 spectra by the normalized LTT 4364 spectrum corrected 
well for telluric features, even though NGC 3105 and LTT 4364 were observed 
on different nights. Division by the LTT 4364 spectrum 
also removes variations in system response over short 
and intermediate wavelength intervals, leaving pseudo continuum-corrected 
spectra that are normalized to the spectral energy distribution (SED) 
of LTT 4364. The final step in the processing was to fit a continuum function to 
each spectrum, and divide each telluric-corrected spectrum by the result.

\LongTables
\begin{deluxetable}{cccccl}
\tablecaption{Spectroscopic targets}
\startdata
\tableline\tableline
ID & RA & Dec & $i'$ & $g'-i'$ & Region\tablenotemark{a}\tablenotemark{b} \\
 & (2000) & (2000) & & & \\
\tableline
1 & 150.236249 &-54.767349 &15.823 &0.955 & F \\
2 & 150.232644 &-54.800179 &18.941 &1.578 & F \\
3 & 150.228896 &-54.806980 &15.656 &0.917 & F \\
4 & 150.224848 &-54.790791 &16.473 &1.212 & F* \\
5 & 150.219455 &-54.798790 &16.305 &1.200 & F \\
6 & 150.215549 &-54.787201 &18.850 &1.443 & F \\
7 & 150.206852 &-54.789371 &18.755 &1.519 & F \\
8 & 150.202804 &-54.808510 &18.731 &1.473 & F \\
9 & 150.199199 &-54.790760 &18.671 &1.517 & S \\
10 & 150.195293 &-54.784309 &20.509 &2.310 & S \\
11 & 150.193750 &-54.784167 &18.576 &2.018 & S \\
12 & 150.191846 &-54.792011 &20.532 &2.058 & S \\
13 & 150.188556 &-54.800850 &20.637 &2.551 & S \\
14 & 150.184951 &-54.813831 &16.347 &1.128 & F \\
15 & 150.181503 &-54.808170 &18.467 &1.452 & F \\
16 & 150.177312 &-54.799061 &20.553 &2.114 & C \\
17 & 150.172648 &-54.784229 &20.553 &2.210 & C \\
18 & 150.166197 &-54.797981 &20.486 &1.974 & C \\
19 & 150.162449 &-54.801311 &20.560 &2.422 & S \\
20 & 150.158844 &-54.792580 &20.499 &2.205 & C \\
21 & 150.153751 &-54.789532 &20.560 &2.113 & C \\
23 \tablenotemark{c} & 150.148501 &-54.798698 &20.572 &2.165 & S \\
24 & 150.145354 &-54.794891 &18.569 &1.484 & S \\
25 & 150.141006 &-54.789539 &20.462 &2.269 & S \\
26 & 150.136042 &-54.795441 &20.473 &2.138 & F \\
27 & 150.135417 &-54.795417 &21.148 &2.566 & F \\
28 & 150.126901 &-54.783588 &16.024 &0.968 & F \\
29 & 150.117602 &-54.784351 &15.980 &1.171 & F* \\
30 & 150.107846 &-54.801510 &15.844 &0.911 & F \\
31 & 150.103812 &-54.774899 &16.240 &1.180 & F* \\
32 & 150.099167 &-54.801750 &15.378 &2.632 & F \\
33 & 150.098405 &-54.801609 &16.113 &1.014 & F \\
34 & 150.094800 &-54.812679 &18.582 &1.333 & F \\
35 & 150.091352 &-54.787239 &15.948 &1.018 & F \\
\tableline
\enddata
\tablenotetext{a}{F = Field, S = Shoulder, C = Center}
\tablenotetext{b}{Mask alignment stars indicated with a `$\ast$'}
\tablenotetext{c}{There is no entry for Star 22 as it was found to be an artifact 
of a bright star.}
\end{deluxetable}

\subsection{SPITZER Imaging}

	Images recorded with the SPITZER IRAC as part of the GLIMPSE survey 
(Benjamin et al. 2003) are used in this study to examine the light 
distribution of NGC 3105 at mid-infrared (MIR) wavelengths (Section 3), and also 
to investigate the MIR photometric properties of the brightest stars. 
Post Basic Calibrated Data (`PBCD') images of NGC 3105 and its surroundings 
were downloaded from the Spitzer Science Archive \footnote[5]
{http://irsa.ipac.caltech.edu/data/SPITZER/GLIMPSE/}. The 
PBCD mosaics have 0.6 arcsec pixel$^{-1}$ sampling, with an 
angular resolution of $\sim 1.7$ arcsec FWHM in [3.6] and [4.5] 
(Fazio et al. 2004). The effective exposure time is 2 seconds per pixel. 

\section{LIGHT PROFILES OF NGC 3105}

	The extremes of the IRAC wavelength coverage are plumbed by the 
[3.6] and [8.0] filters, which probe different sources of astrophysical emission. 
The signal at the short wavelength end of the IRAC coverage is predominantly 
stellar in origin, while the signal in [8.0] may contain a large contribution from 
interstellar PAH emission, which has a strong feature near $7.6\mu$m and so 
falls in the [8.0] pass band. The portions of the [3.6] and [8.0] images 
that sample the region near NGC 3105 are shown in Figure 1. While not obvious from 
Figure 1, there is patchy emission in and around NGC 3105 in 
the [8.0] PBCD mosaic, with locally strong emission East of the cluster. 

	Efforts to trace the light profile of NGC 3105 
outside of its central regions are hindered by stochastic effects and 
contamination from field stars. The S/N ratio of the cluster light profile 
can be boosted by azimuthally averaging the signal after assuming a 
shape and orientation for the cluster. The IRAC images of NGC 3105 were binned 
$12.6 \times 12.6$ arcsec to suppress individual bright stars, and the results were 
divided into 24 azimuthal zones centered on the cluster. The azimuthal slices were 
aligned, and the median signal at each pixel location found. 
Pixels that contain field stars -- which are randomly distributed 
-- will be rejected when taking the median, while signal from the cluster light will 
reinforce constructively. The light profile was then extracted from the median 
pixel values. This procedure assumes circular isophotes. 

	The [3.6] and [8.0] light profiles are compared 
in the top panel of Figure 2. The cluster center was assumed to be at ($\delta$, RA) = 
(--54:47:17.6, 10:00:40.1; E2000), which is near the middle 
of an asterism defined by four bright, centrally-concentrated stars. 
The PBCD images cover $\sim 0.9$ degrees$^2$, allowing background light 
levels to be measured at large distances from the cluster. Still, 
there is uncertainty in the [8.0] background level because of 
wide-spread, non-uniform emission throughout the sampled area.

	Light from NGC 3105 is traced out to 
at least 100 arcsec in [3.6], and the FWHM of the light profile is $\sim 50$ arcsec. 
The localized peak in the [3.6] profile near $r \sim 25$ arcsec is residual signal 
from the bright stars that surround the cluster center (Figure 1). 
It can be seen in Figure 1 that NGC 3105 is not as obvious in [8.0] 
as it is in $i'$ or [3.6]. In fact, the [8.0] light profile in 
Figure 2 is much flatter than the [3.6] profile, with a FWHM $\sim 250$ 
arcsec. The shape of the [8.0] profile follows that of 
the [3.6] profile at radii $< 30$ arcsec, suggesting that 
the light in [8.0] at these radii is dominated by stars in NGC 3105, rather than 
PAH emission. 

\begin{figure}
\figurenum{2}
\epsscale{1.2}
\plotone{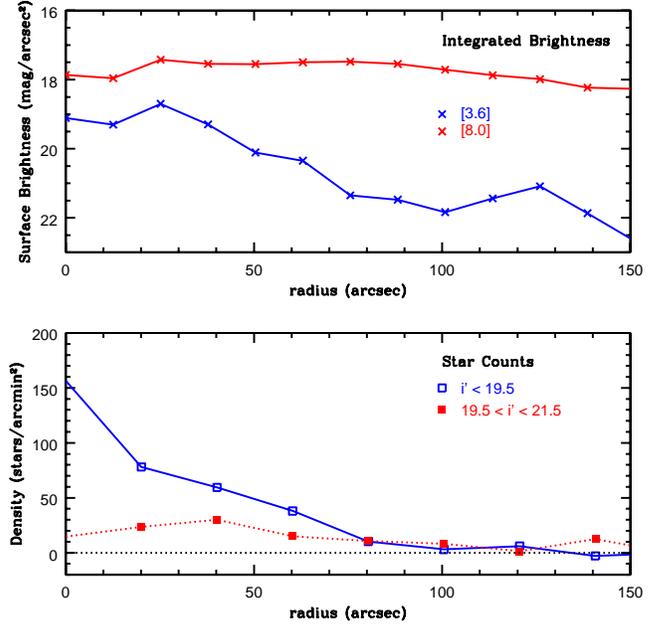}
\caption{(Top panel:) The [3.6] (blue) and [8.0] (red) light profiles of NGC 3105. 
Background sky levels were measured at large angular offsets 
from the cluster center. The peak in surface brightness near 25 arcsec 
is an artifact of the constellation of bright stars that 
surround the cluster center. The [8.0] light profile is 
flatter than the [3.6] light profile, and this is attributed to 
interstellar PAH emission, although cluster stars also contribute to the signal. 
(Lower panel:) Number counts of stars with $i' < 19.5$ (open blue squares) and 
$i'$ between 19.5 and 21.5 (filled red squares) made in 20 arcsec wide annuli. 
Contamination from field stars has been corrected by subtracting star counts 
in Field 2 (Section 5.1), which samples distances $> 177$ arcsec 
from the cluster center. Stars with $i' > 19.5$ are more centrally concentrated than
the stars in the fainter sample.}
\end{figure}

	Images in [4.5] and [5.8] were also recorded as part of the GLIMPSE survey. 
The NGC 3105 light profile in [4.5] -- which is not shown here -- is almost identical 
to the [3.6] profile. The light distribution in [5.8] -- also not shown here -- 
has a width that is intermediate between those in [3.6] and [8.0]. There is a PAH 
emission feature near $6.2\mu$m that falls within the [5.8] pass band.

	Light profiles were also constructed from images recorded as part of 
the Wide-field Infrared Survey Explorer (WISE; Wright et al. 2010) 
Allsky survey. Images in the W1 ($\lambda_{cen} \sim 3.4\mu$m) and W3 
($\lambda_{cen} \sim 12\mu$m) filters were downloaded 
from the IPAC archive \footnote[6]{http://irsa.ipac.caltech.edu/Missions/wise.html}. 
The images were azimuthally smoothed using the procedure described above, 
and the light profiles extracted. As with the IRAC data, the W1 and W3 light profiles 
are very different, with the W3 profile being much broader than the W1 profile. 
The W3 bandpass samples PAH emission features.

\section{PHOTOMETRIC MEASUREMENTS}

	Stellar photometry was done with the point 
spread function (PSF)-fitting routine ALLSTAR (Stetson \& Harris 1988). 
Tasks in the DAOPHOT package (Stetson 1987) were used to generate the source 
catalogues, preliminary brightnesses, and PSFs that serve as input to ALLSTAR. The PSFs 
were constructed from stars that are bright, isolated, and unsaturated. Contamination 
from faint stars close to the PSF stars was removed iteratively, using progressively 
improved versions of the PSF to subtract out these objects.

	The photometric measurements for all but the brightest stars in the GMOS 
images were made by fitting the PSF within the FWHM of the stellar profiles. 
However, the bright MS and evolved stars in NGC 3105 that 
are important for estimating age are saturated in the short 
$i'$ exposures. While photoelectric magnitudes have been published for the 
brightest stars in and around NGC 3105 (Moffat \& Fitzgerald 1974; Fitzgerald et al. 
1977), these are not in the photometric system used here. To recover the brightnesses 
of these stars, ALLSTAR was run a second time on the 
short exposure $i'$ image, but this time with the PSF fit over a much larger portion 
of the stellar profile than just the central regions. The 
central, saturated parts of the stellar profiles are rejected in the fitting 
process, so that profile fitting is done only in the unsaturated outer regions 
of the PSF. 

	The IRAC photometry was calibrated using zeropoints from Reach et al. (2005).
Zeropoints for the GMOS photometry were obtained from observations of standard stars 
that were recorded on January 1 2014. The standard stars are 
in the 075944--59550 field, and their magnitudes were 
taken from the Southern Standard Stars for the u'g'r'i'z' System 
website \footnote[7]{http://www-star.fnal.gov/}. The zeropoints obtained from these 
observations fall within the range of values archived for this instrument 
\footnote[8]{http://www.gemini.edu/sciops/instruments/performance-monitoring/data-products/gmos-n-and-s/photometric-zero-points}. 

\section{COLOR-MAGNITUDE DIAGRAMS}

\subsection{GMOS CMD}

	$(i', g'-i')$ CMDs of the area in and around NGC 3105 are shown in Figure 3. 
The photometry for objects with $i' < 17.5$ is from the short exposure images, 
while the photometry of fainter objects is from the long exposure images. 
The photometric measurements for stars with $i' > 12.5$, which are saturated in 
the short $i'$ exposures, were made by fitting the PSF to the wings of the 
stellar profiles (Section 4). The general appearance of the bright end of the CMD 
is similar to that in previously published CMDs of this cluster.

\begin{figure}
\figurenum{3}
\epsscale{1.2}
\plotone{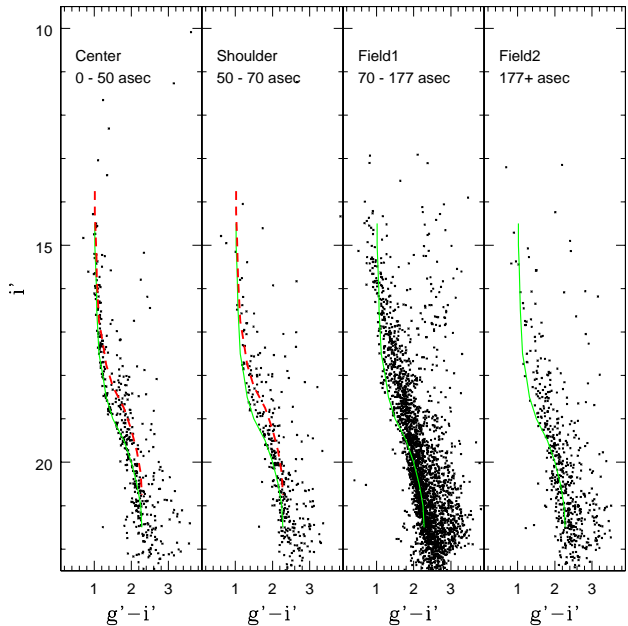}
\caption{The $(i', g'-i')$ CMDs of radial intervals centered on NGC 3105, 
with boundaries defined in part using the [3.6] light profile (see text).
The solid green line is a hand drawn fit to the cluster sequence in the Center field, 
while the dashed red line is this sequence shifted to brighter values by 
0.75 magnitudes to mark the expected location of unresolved equal 
mass binaries. These are the deepest visible/red CMDs yet published for NGC 3105, and 
the cluster sequence can be traced to $i' > 21$ ($g' \geq 23.5$) in the 
Center and Shoulder CMDs. The stellar content of Field 2 is used throughout the 
paper to assess contamination from non-cluster objects.}
\end{figure}

	CMDs are shown for stars in four different regions, with the 
boundaries defined in part using the [3.6] light profile in Figure 2. 
The outer radius of the Center region is at the HWHM of the 
[3.6] light profile, while the outer radius of the Shoulder region is 
where the [3.6] light profile flattens. The 
CMDs in the two remaining panels sample areas that are dominated by field stars, 
although a modest cluster population is likely present in Field 1 (Section 6). 
Field 2 was selected to sample the corners of the GMOS science field, which are the 
areas on the GMOS detector that are furthest from the cluster center. 
For the remainder of the paper it is assumed that Field 2 is free of cluster 
stars, and so this area is used to monitor background/foreground contamination.

	The radial frequency of cluster stars is investigated 
in the lower part of Figure 2, where the radial distributions of objects with $i' 
\leq 19.5$ and $i'$ between 19.5 and 21.5, counted in 20 arcsec wide annuli, 
are compared. Star counts made in Field 2 were subtracted from the number 
counts to correct for non-cluster stars. There is a clear tendency for objects 
in the brighter sample to be concentrated within $\sim 80$ arcsec of the cluster 
center. There is also a tendency for stars in the faint sample 
to be located within $\sim 100$ arcsec of the cluster center, although 
there is not a central cusp. For comparison, Moffat \& Fitzgerald (1974) found that a 
significant number of cluster stars with $B \leq 17.2$ are detected out to 
distances of 50 - 80 arcsec from the cluster center, although they 
also argue that some cluster members -- identified as 
such from their photometric properties -- may occur out to radii 
of $\sim 120+$ arcsec. The star counts shown in the lower panel of Figure 2 
suggest that cluster stars are not present in large numbers in Fields 1 and 2.

	The green line in Figure 3 is a hand-drawn representation of the 
cluster sequence in the Center CMD. The cluster sequence can be traced 
to $i' \sim 21$ in the Center and Shoulder regions, which corresponds 
to $g' \sim 23.5$, or $V \sim 22.5 - 23.0$. For comparison, the faint 
limits of the Paunzen et al. (2005) and Sagar et al. (2001) CMDs 
are $V \sim 18$. These are thus the deepest (by many magnitudes) 
photometric measurements yet obtained of NGC 3105 at visible/red wavelengths. 
Below, it is shown that observations at this depth can provide 
constraints on the age of NGC 3105, and also sample a magnitude range where 
signatures of mass segregation become evident in the cluster (Section 6). 

	Stars with $i' < 12.5$ are concentrated in the Center CMD, and 
the blue stars with $i'$ between 11.5 and 
14 in the Center CMD are likely upper MS or post-MS objects (see below). 
There are also two red stars with $i' < 11.2$. If -- as suggested by their 
location near the cluster center -- these are cluster members then they 
are red supergiants (RSGs).

	The Field 1 and Field 2 CMDs are dominated by richly populated sequences with 
ridgelines that fall redward of the green cluster fiducial. Fore/background 
stars in the Field 1 CMD form a near-linear sequence that intersects 
the NGC 3105 fiducial near $i' = 16$ and $i' = 21$. Contamination from 
non-cluster stars is substantial when $i' > 20$, 
although number counts discussed below suggest that 
solar and sub-solar mass stars associated with NGC 3105 
may be present in Field 1. Still, the locus of 
stars with $i' > 20$ in the Field 1 CMD falls redward of the 
sequence defined from the Center CMD. 

	The dispersion in the cluster sequence near $i' \sim 16$ 
is $\sigma (g'-i')  = \pm 0.06$ magnitudes. If due only 
to variations in line of sight extinction then this 
corresponds to a $\sim \pm 0.04$ magnitude dispersion in E(B--V). While the narrow 
nature of the cluster sequence might also suggest that the incidence of unresolved 
binaries is small, this may not be the case. The dashed line 
in Figure 3 shows the cluster fiducial shifted upwards by 0.75 magnitudes 
to simulate the location of unresolved binaries with a mass 
ratio = 1. If present in large numbers then it can be seen from Figure 3 
that unresolved equal-mass binaries will form a conspicuous sequence 
in the cluster CMD between $i'$ of 17 and 21.

	While there are stars in the Center and Shoulder region CMDs that are close 
to the binary sequence, there is a significant contribution from field stars. 
The number of equal mass binaries in NGC 3105 have been estimated by counting 
objects within $\pm 0.1$ magnitudes in $g'-i'$ of the binary sequence 
with $i'$ between 18 and 19. There are 29 sources in this part of the 
Center and Shoulder region CMDs, whereas the Field 2 CMD contains 
18 sources. The mean densities on the sky are then $6.8 \pm 1.3$ arcmin$^{-2}$ 
in the Center and Shoulder regions and $3.2 \pm 0.8$ arcmin$^{-2}$ in Field 2. 
Assuming that Field 2 contains no cluster stars then the 
difference in projected densities is significant at the $2.4 \sigma$ 
level, suggesting that NGC 3105 harbors a population of unresolved, 
equal-mass binaries. Based on the number of stars within $\pm 0.1$ magnitudes in 
$g'-i'$ of the main sequence between $i$ of 18.75 and 19.75, we estimate a 
binary frequency of $0.5 \pm 0.2$ for equal mass systems. We caution 
that single stars in NGC 3105 with higher than average extinction may have photometric 
properties that have been shifted into the area of the CMD that is 
occupied by equal mass binaries.

	The combined CMD of the Center and Shoulder regions is shown in 
Figures 4 and 5, where comparisons are made with Padova (Bressan et al. 2012) Z = 0.020 
(Figure 4) and Z=0.008 (Figure 5) isochrones. The models from which these 
isochrones were constructed include PMS evolution. The isochrones were positioned 
to match (1) the blue edge of the upper MS, which provides 
leverage on reddening, and (2) the bend in the CMD that occurs 
between $i' = 18$ and 20, which provides constraints on distance. The reddenings 
and distance moduli that are found by comparing the isochrones with the observations 
are listed in each figure. The reddenings found from the Z=0.008 and 
Z=0.020 isochrones are the same, as the intrinsic colors of hot stars are 
not sensitive to metallicity. However, the distances 
differ, with d $ = 7.9^{+0.4}_{-0.3}$ kpc for Z=0.020, and $6.6 \pm 0.3$ 
kpc for Z=0.008. A $\pm 0.1$ magnitude uncertainty in the distance modulus has 
been assumed, which was found by perturbing the adopted distance 
modulus and assessing the agreement between the models and observations.

\begin{figure}
\figurenum{4}
\epsscale{1.2}
\plotone{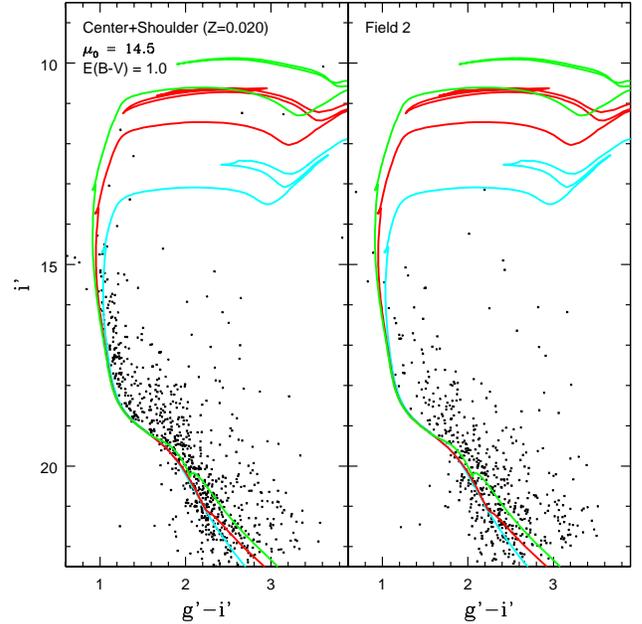}
\caption{Comparisons with Z=0.020 isochrones from Bressan et al. (2012). 
Sequences with ages 20 Myr (green), 32 Myr (red), and 71 Myr (cyan) are shown. 
The distance modulus and reddening that result from matching the 
isochrones to the blue envelope of the cluster sequence are listed in the left hand 
panel. These models rule out an age as young as 20 Myr 
based on the color of the cluster sequence at $i' > 20$, and this is consistent with 
the photometric properties of what are likely the most evolved cluster members.}
\end{figure}

\begin{figure}
\figurenum{5}
\epsscale{1.2}
\plotone{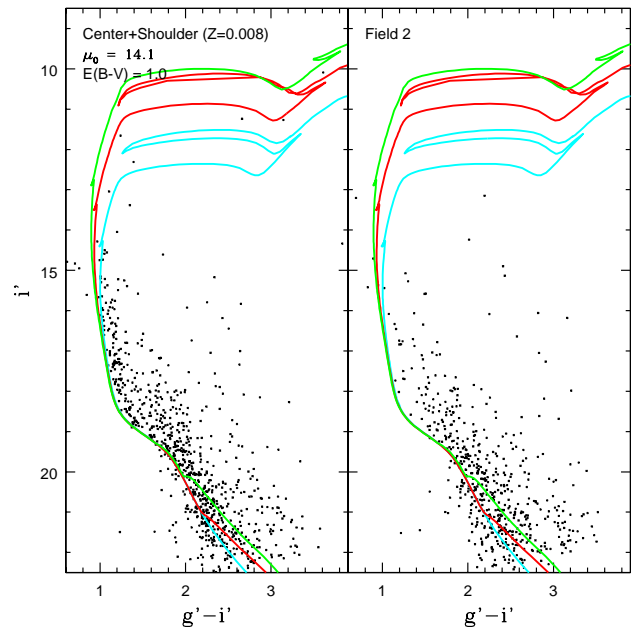}
\caption{Same as Figure 4, but showing comparisons with Z=0.008 isochrones. Isochrones 
with this metallicity and an age $\sim 32$ Myr match the three brightest evolved stars, 
while stars with $i'$ between 12 and 14 scatter about the 71 Myr model. 
The CMD at $i' > 20$ favors an age older than 20 Myr.}
\end{figure}

	The 32 Myr Z=0.020 model best matches the photometric 
properties of stars with $i' > 13$, although the isochrones 
extend far redward of the three brightest objects in the left 
hand panel that are presumably the most evolved members of the cluster. 
As for the faint end of the CMD, the 20 and 32 Myr Z=0.020 models
more-or-less bracket the main body of stars with $i' > 20$. The 
comparisons with the Z=0.020 isochrones thus point to an age $> 20$ Myr.

	NGC 3105 is in the outer regions of the Galaxy, and so it likely has a 
sub-solar metallicity. It is thus worth noting that the location of the three candidate 
RSGs on the CMD fall near the redward extent of the Z=0.008 isochrones. 
Still, stars with $i'$ between 12 and 14 tend to scatter near 
the 71 Myr isochrone. The 20 Myr sequence falls redward of the main 
concentration of stars at $i' > 20$.

	The color distributions in the Center$+$Shoulder and Field 2 CMDs 
in $\pm 0.25$ magnitude intervals centered at $i' = 19.5, 20.0,$ and 20.5 
are compared in Figure 6. A 0.2 magnitude binning in $g'-i'$ was found to 
deliver statistically significant numbers of stars in each 
color interval. The results of subtracting the Field 2 color 
distribution from that of the Center$+$Shoulder distribution are shown in 
the right hand column.

\begin{figure}
\figurenum{6}
\epsscale{1.2}
\plotone{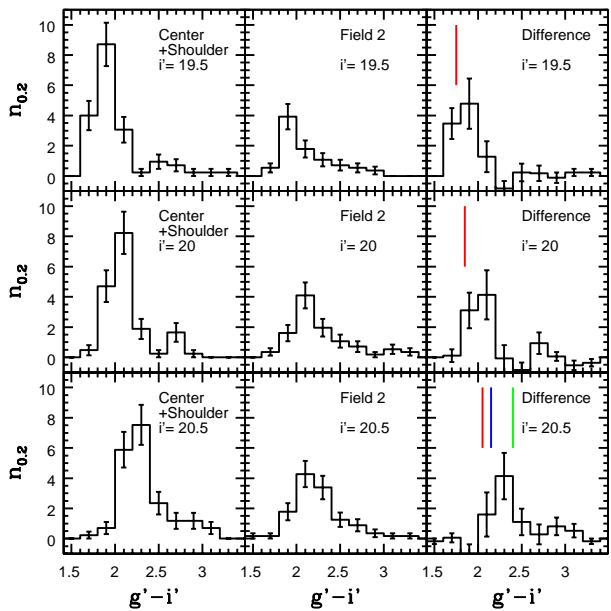}
\caption{$g'-i'$ distributions in three $i'$ intervals. n$_{0.2}$ is the number of 
objects per arcmin$^2$ in $\pm 0.25$ magnitude $i'$ 
intervals with 0.2 magnitude binning in $g'-i'$. 
The distributions in the combined Center$+$Shoulder region are shown in 
the left hand columns while the color distributions in Field 2, which samples the 
largest radial distances from the cluster center, are shown in the middle column. The 
differences between the Center$+$Shoulder and Field 2 
distributions are shown in the right hand panels. The red line indicates 
the color of the Z=0.008 32 Myr model at each brightness, while the green and 
blue lines show the color of 20 and 22 Myr models at $i' = 20.5$ -- the colors of 
the 20 and 22 Myr isochrones are not indicated in the $i' = 19.5$ 
and 20.0 panels as they are the same as the 32 Myr model. 
These comparisons favor an age for NGC 3105 that is older than 20 Myr.}
\end{figure}

	Contamination from non-cluster stars 
accounts for $\sim 50\%$ of the objects with $g'-i'$ between 2 and 2.5. 
This level of contamination notwithstanding, there remains a statistically 
significant sample of objects after subtracting the Field 2 color distribution. 
The colors of 20, 22 and 32 Myr Z=0.008 sequences 
from Bressan et al. (2012) are shown in the right hand column of Figure 6. 
All three sequences have the same colors at $i' = 19.5$ and $i' = 20$, and 
so only the 32 Myr model is shown in those panels. The situation changes 
at $i' = 20.5$, where the majority of stars fall between the 20 and 32 Myr models. 
It can be seen from the color of the 22 Myr isochrone that $g'-i'$ changes 
quickly with age at this brightness. The persistent tendency 
for the 32 Myr sequence to fall blueward of the 
peak in the color distribution at all magnitudes is due to the matching 
of the isochrones to the blue edge of the cluster sequence in the CMD. 

	The comparisons in Figure 6 suggest that an age as young as 20 Myr for NGC 
3105 can be ruled out, but that an age that is only 2 Myr older would be consistent 
with the color of the cluster sequence. We note that previous studies have encountered 
difficulties reproducing the colors of PMS stars at visible and red wavelengths (e.g. 
Lyra et al. 2006; Bell et al. 2012). Davidge (2015) found an offset 
$\Delta (g'-i') \sim 0.4$ magnitudes between the 
observed and model PMS sequence in the $\sim 20$ Myr cluster NGC 2401, in the 
sense that the cluster sequence is redder than models from 
Bressan et al. (2012) and Siess et al. (2000). Still, unless there are 
systematic differences between the properties of PMS stars in NGC 2401 and 
NGC 3105 (aside from age) then the $g'-i'$ color of the faint end of the NGC 3105 
CMD suggests that the cluster has an age $> 20$ Myr. Folding in the properties 
of the three brightest stars, which are assumed to belong 
to NGC 3105 and are matched by the 32 Myr isochrones, and 
considering that the brightest MS stars match the 71 Myr models, we conclude that 
NGC 3105 has an age of at least 32 Myr. 

\subsection{SPITZER CMDs}

	Some of the brightest members of NGC 3105 and the surrounding field 
are shrouded by circumstellar envelopes. One tracer of circumstellar 
material -- H$\alpha$ emission -- is detected among bright early-type stars in 
and around NGC 3105 (Moffat \& FitzGerald 1974; FitzGerald et al. 1977), 
and additional emission line sources are discussed in 
Section 7. Some of the brightest stars in NGC 3105 
also may have Br$\gamma$ in emission (e.g. Figure 6 of Davidge 2014).

	If there are thick dust envelopes around some of the stars in 
NGC 3105 then these objects might also show excess MIR emission. 
The $([4.5],[3.6]-[4.5])$ and $([8.0],[3.6]-[8.0])$ CMDs of objects in and 
around NGC 3105 are shown in Figures 7 and 8. These CMDs cover the same radial 
intervals as in Figure 3. Only post-MS and bright MS stars in NGC 3105 
are sampled by the GLIMPSE observations. 

\begin{figure}
\figurenum{7}
\epsscale{1.2}
\plotone{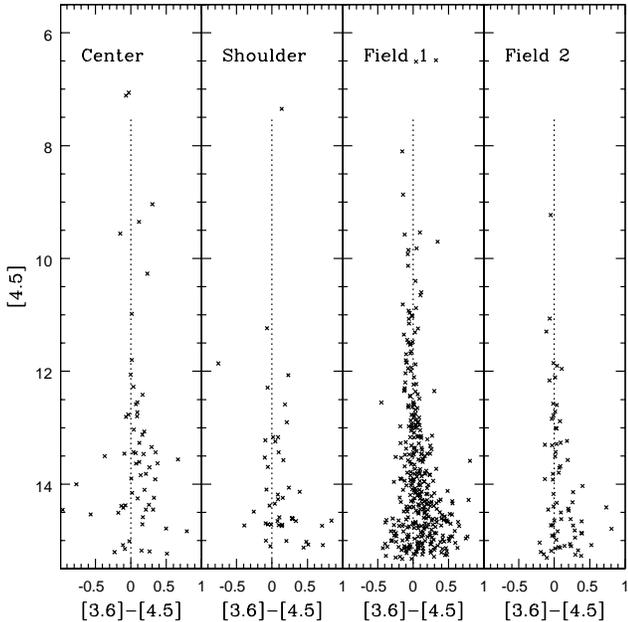}
\caption{$([4.5], [3.6]-[4.5])$ CMDs of NGC 3105 and the surrounding area. 
The dotted line in each panel indicates [3.6]--[4.5] = 0, which is 
the approximate color of photospheric light. 
Some of the stars in the Center region with $i'$ between 9 and 11 have redder 
[3.6]--[4.5] colors than the majority of objects, 
and this is likely due to emission from circumstellar material.}
\end{figure}

\begin{figure}
\figurenum{8}
\epsscale{1.2}
\plotone{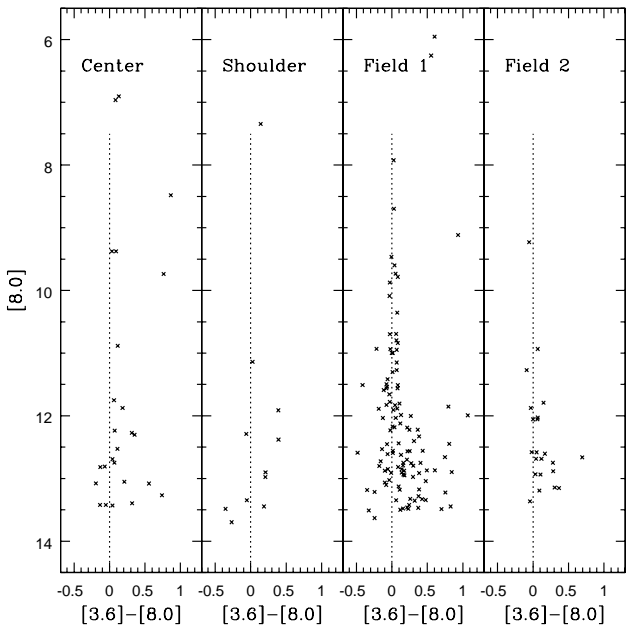}
\caption{Same as Figure 7, but showing $([8.0], [3.6]-[8.0])$ CMDs. 
The bright stars near the cluster center that have red [3.6]--[8.0] 
colors are the same ones that have red [3.6]--[4.5] colors in Figure 7.}
\end{figure}

	The prominent vertical plume in the SPITZER CMDs is populated by objects that 
have a range of effective temperatures. While having very different spectral-types, 
these objects have similar [3.6]--[4.5] and [3.6]--[8.0] colors (roughly 0 in each 
case) because the Rayleigh-Jeans tail of their SEDs is 
sampled at these wavelengths. Still, there are two bright 
objects in the Center field that have redder [3.6]--[8.0] colors than the majority of 
cluster (and field) members. These are among the brightest cluster members, and 
their location on the GMOS CMD indicates that they are evolving near the MSTO -- 
hence, they are likely among the most massive stars in NGC 3105. The brightest of 
these is Star 7 in the compilations of Moffat \& Fitzgerald (1974) and 
FitzGerald et al. (1977), which they classify as a Be star. A modest number of 
sources with red [3.6]--[4.5] and [3.6]--[8.0] colors are also seen in 
Field 1, and in Section 7 it is shown that emission line stars are also detected 
in this part of the sky.

\section{THE LUMINOSITY FUNCTION AND SPATIAL DISTRIBUTION OF STARS}

	Cluster LFs contain encoded information about the mass function (MF) 
of the component stars. The $i'$ LFs of the Center and Shoulder regions are shown 
in Figure 9. Only stars that were detected in both filters were counted.

\begin{figure}
\figurenum{9}
\epsscale{1.2}
\plotone{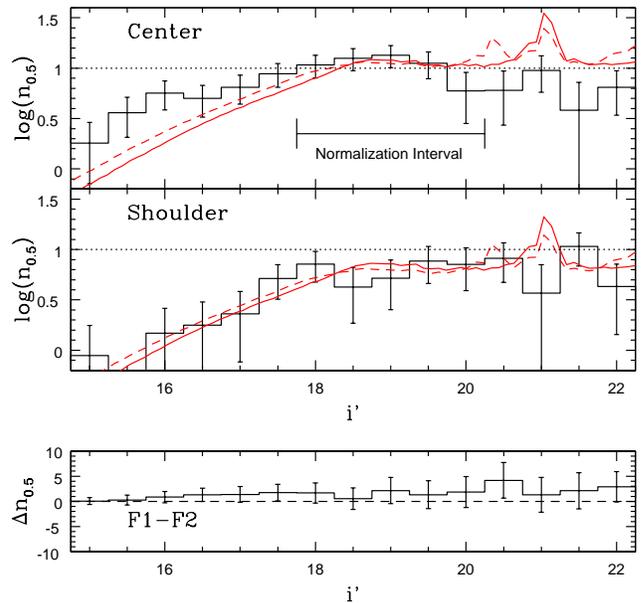}
\caption{ (Top two panels): $i'$ LFs of the Center and Shoulder 
fields. n$_{0.5}$ is the number of stars per square arcmin per 
0.5 magnitude in $i'$, corrected statistically for non-cluster objects by subtracting  
the Field 2 LF after scaling for differences in sky coverage. The error bars show 
counting statistics. The dotted line in each panel marks log(n$_{0.5}) = 1$. While 
the number density of stars with $i' > 19.5$ is more-or-less similar in the two 
regions, the number of MS stars with $i' < 18$ clearly drops with increasing radius -- 
the ratio of bright to faint stars in the Center region is higher than in the 
Shoulder region. The red line is a model LF generated from the Bressan et al. (2012) 
isochrones for single stars (i.e. no binaries), assuming an age of 32 Myr, Z=0.008, 
and the distance modulus and E(B--V) used in Figure 5. A Chabrier (2003) lognormal 
MF has been adopted, and the LF has been scaled to match the observations 
in the $i'$ interval that is indicated. The dashed red line shows a 
model LF that includes equal mass binaries with a 0.5 binary fraction. 
(Lower panel:) Difference in number counts between Field 1 and Field 2. 
There is a systematic tendency for Field 1 to contain more objects than Field 2, 
suggesting that some stars associated with NGC 3105 may be present in Field 1.}
\end{figure}

	Stars associated with NGC 3105 may be 
present in Field 1, and evidence to support this is presented in the lower panel of 
Figure 9, where the difference between the Field 1 and 2 LFs is shown. 
While there are no statistically significant differences 
between the two LFs in any one $i'$ bin, the counts in Field 1 are systematically 
higher than those in Field 2 at all magnitudes, suggesting that objects 
associated with NGC 3105 are present in Field 1. A modest excess numbers of stars in 
Field 1 when compared with Field 2 is also consistent with the [3.6] surface 
brightness profile in Figure 2. Field 2 was thus adopted as the background field, and 
a statistical correction for field stars was applied to the LFs in 
Figure 9 by subtracting the Field 2 LF. 

	The projected density of objects 
with $i' \geq 19.5$ is similar in the Center and Shoulder regions. 
When averaged over tens of arcsec, objects with $i'$ between 
19.5 and 22.0 are thus more-or-less uniformly distributed within 70 arcsec of the 
cluster center. In contrast, the number of stars with $i' \leq 19$ in the Center region 
is $\sim 0.4 - 0.5$ dex (i.e. a factor of $\sim 3\times$) higher than in the 
Shoulder region. These differences in the distribution of bright and faint 
stars point to mass segregation in NGC 3105.

	To further explore the nature of any radial variation in stellar 
content, model LFs based on the Z=0.008 Padova 32 Myr isochrone were 
constructed, and these are shown in Figure 9. The models assume a Chabrier (2003) 
lognormal mass function, and have been normalized to match the number counts in 
the magnitude interval indicated in the top panel. The distance modulus and 
reddening used in Figure 5 have been adopted for this comparison. 

	The solid line in Figure 9 shows a model LF that assumes no binaries, 
whereas the dashed line shows a model that includes equal mass binaries with a 
binary fraction 0.5, as estimated in Section 5. As 
expected, the model that includes binaries falls above the non-binary 
model at the bright end. A model LF that assumes the same binary fraction but a 
Chabrier mass distribution for companions (i.e. mass ratios different from unity)
would fall midway between the equal mass and non-binary models. 

	The bump in the single star model LF in Figure 9 near $i' = 21$ is due to 
the MSCO, which can be used to estimate age if the distance and 
reddening of a young cluster are known. The amplitude of the MSCO is strongest 
in very young clusters, and weakens with increasing age. The MSCO will be difficult 
to detect in NGC 3105 given the statistical errors in the number counts, and 
there is no corresponding feature in the cluster LF near this magnitude. 

	The models in Figure 9 match the observations in the Shoulder region, 
suggesting that the isolated stars and dominant members of binary systems 
in that part of NGC 3105 can be characterized by a Chabrier MF, at least 
as faint as $i' = 22$ ($\sim 0.7 - 0.8$M$_{\odot}$). In contrast, there is a systematic 
excess in the Center region LF when compared with the models in the magnitude interval 
$i' < 18$ ($> 1.7$M$_{\odot}$). The difference between the models and observations 
in the Center region would become larger yet if the models were normalized at fainter 
magnitudes than shown in Figure 9. The models indicate that binarity affects the 
number counts at the bright end of the LF by no more than $\sim 0.2$ dex, and so 
it is unlikely that binarity can explain the central excess of bright objects.

\section{STELLAR SPECTRA}

	The locations and photometric properties of stars for which spectra were 
obtained are listed in Table 3, while the $(i', g'-i')$ CMD of these objects, 
with Z=0.008 isochrones from Bressan et al. (2012) over-plotted, is shown in Figure 10. 
The targeted stars sample three narrow magnitude intervals to facilitate star-to-star 
comparisons at different points on the cluster CMD. Brightness-related biases 
are reduced because the stars in each group have comparable magnitudes.
The two objects with unusually red colors near $i' \sim 15.5$ 
and 18.5 were observed because they could be placed in a slit that contained a primary 
spectroscopic target. The faintest star in the sample, with $i' \sim 21.0$, was 
also observed for this reason.

\begin{figure}
\figurenum{10}
\epsscale{1.2}
\plotone{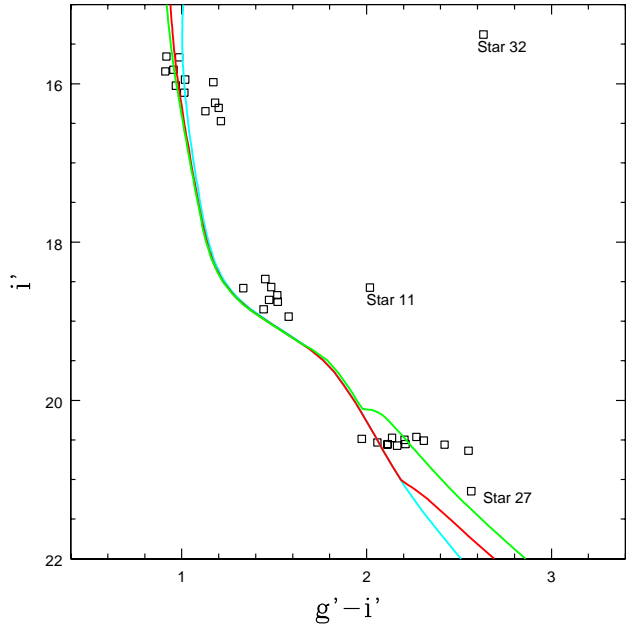}
\caption{CMD of stars for which spectra were recorded. 
The target stars sample three parts of the CMD: (1) the upper MS (A--B stars), 
(2) the middle part of the MS (A--F stars), and (3) the magnitude 
where PMS stars might be expected based on past age estimates (likely 
G dwarfs). Stars 11, 27, and 32 were observed because they could 
be placed in the same slit as a high priority target. Z=0.008 
isochrones for ages 20 Myr (green), 32 Myr (red), and 72 Myr (cyan) 
are also shown.}
\end{figure}

	Cluster members in the brightest group sample the upper MS. Members of 
NGC 3105 in this group will have M$_V \sim 0$, and so are 
expected to have A -- B spectral types. Based on previous spectroscopic studies 
of stars near NGC 3105 (Moffat \& FitzGerald 1974; FitzGerald 
et al. 1977), the spectra of some of these might contain emission features.
Early-type stars are also of interest for assessing interstellar absorption as 
there are large wavelength intervals that are free of strong photospheric lines. 

	The stars with M$_{i'} \sim 18.5$ sample objects midway down the MS in 
NGC 3105 with M$_V \sim 2.5$. Stars with this M$_V$ have A -- 
F spectral-types, and these spectra may contain detectable metallic 
absorption lines. The faintest group, with M$_{i'} \sim 20.5$, 
samples stars in NGC 3105 that have M$_V \sim 5.5$. 
Cluster members at the brightness will be G dwarfs if they are on the MS.

\subsection{The Bright Star Sample}

	The bright sample can be sub-divided into two groups - one with 
$g'-i' < 1.1$ that hugs the isochrone sequence in Figure 10, 
and a second that has redder colors. Spectra centered on Na D 
and H$\alpha$ of stars in the blue and red sub-samples 
are shown in Figures 11 and 12. The mask alignment stars, 
the spectra of which were recorded through a square aperture rather than 
a slit, are indicated with an asterisk. All three alignment 
stars are in the red sub-sample. None of the stars in the 
bright sample are in the Center or Shoulder regions, as primacy for spectroscopic 
targeting in those areas was assigned to fainter objects.

\begin{figure}
\figurenum{11}
\epsscale{1.2}
\plotone{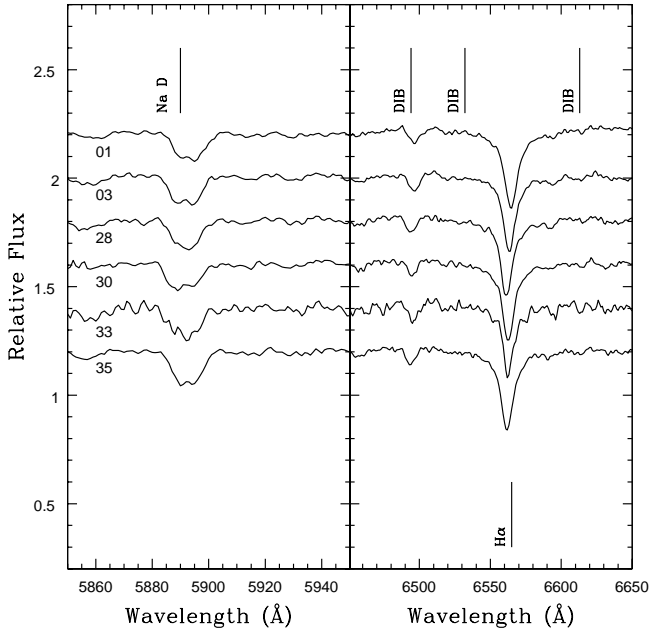}
\caption{Spectra of stars in the bright sample that have $g'-i' 
< 1.1$. Wavelengths centered near Na D and H$\alpha$ are shown. 
DIBs that have mean EWs $> 0.1$ \AA\ in the Diffuse Interstellar Band Catalogue 
(http://leonid.arc.nasa.gov/DIBcatalog.html) are indicated.} 
\end{figure}

\begin{figure}
\figurenum{12}
\epsscale{1.2}
\plotone{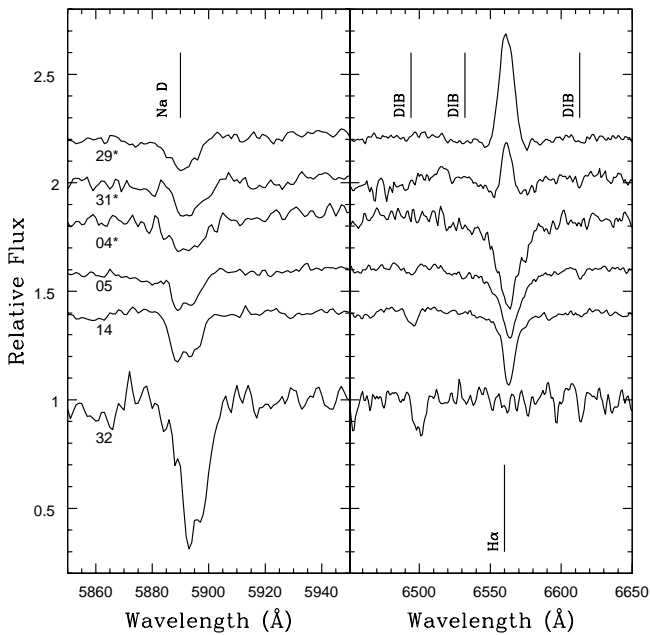}
\caption{Same as Figure 11 but showing spectra of stars in the bright sample with 
$g'-i' > 1.1$. Alignment stars are indicated with 
an asterick next to their identification \#. The stars with 
H$\alpha$ emission are placed at the top for display purposes. 
There is noticeable star-to-star dispersion in the depths of Na D and the DIB at 
6494\AA. Star 32 has spectroscopic properties that are very different from those 
of the other bright objects, and it is likely a foreground K dwarf (Section 7.3).} 
\end{figure}

	There are only modest star-to-star differences
between the spectra in Figure 11. However, this is not the case 
among the spectra in Figure 12. Stars 29 and 31 are alignment stars, 
and have H$\alpha$ in emission, suggesting that they are Be stars. 
Such stars are associated with systems that have ages up to a few tens 
of millions of years, and Fitzgerald et al. (1977) identify three Be stars within 
5 arcmin of the center of NGC 3105. Star 29 in the current sample is likely star 78 
in the Moffat \& FitzGerald (1974) study, although those authors 
did not assign a spectral type for that star. As alignment objects, Stars 29 and 
31 were selected to have a location in the focal plane that did not 
interfere with the spectra of other objects -- that the spectra 
of both show H$\alpha$ in emission provides further evidence that Be stars are 
common in the area around NGC 3105. While Stars 29 and 31 have 
relatively red $g'-i'$ colors, the IRAC photometry 
indicates that neither shows excess IR emission at wavelengths $\leq 8\mu$m, 
although both are near the faint limit of the [8.0] 
GLIMPSE measurements. The absence of an IR excess would suggest that they 
are not of type B[e] (e.g. Allen \& Swings 1976).

	Star 32, which has the reddest $g'-i'$ color in Figure 
10, has a spectrum that differs from that of the other members of the bright group. 
Na D in the Star 32 spectrum is much deeper than in the other spectra, and 
H$\alpha$ is abscent. The EW of Na D in the Star 32 spectrum is $6.5 \pm 0.7$ \AA, 
and this is consistent with that of a mid-K dwarf (e.g. Diaz et al. 2007). If 
Star 32 has a spectral-type K5V (Section 7.3) then $(g'-i')_0 \sim 1.2$ and so 
E(g'--i') $\sim 1.4$. This corresponds to E(B--V) $\sim 0.8$ using the R=3.1 
reddening law of Cardelli et al. (1989). This is marginally lower than the 
reddening of NGC 3105, and is consistent with Star 32 being in the foreground.

	With the exception of Star 32, the stars in the bright sample have 
early spectral types. The H$\alpha$ absorption line profiles tend to be sharper 
than expected from MS stars, and are similar to those in the spectra of 
evolved B stars. This is demonstrated in Figure 13, where the mean spectrum of the 
stars in the blue sub-sample is compared with spectra of B7V (HD 43153)and B7III 
(HD 35497) stars from the Le Borgne et al. (2003) library. Spectra in the Le Borgne et 
al. (2003) compilation have a resolution $\sim 2000$, and so the spectra have 
been smoothed with a Gaussian to match the resolution of the GMOS spectra. 
The H$\alpha$ line profile in the B7V spectrum is clearly broader than in the 
B7III and mean GMOS spectra.

\begin{figure}
\figurenum{13}
\epsscale{1.2}
\plotone{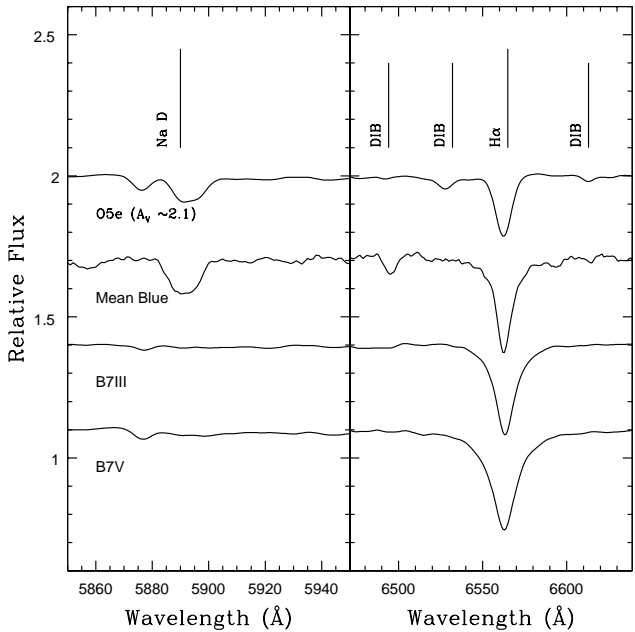}
\caption{Mean spectrum of stars in the bright sample that 
have $g'-i' < 1.1$. Spectra from the Le Borgne 
et al. (2003) library that have been smoothed to match the resolution 
of the GMOS spectra are also shown. Na D is clearly visible in the spectrum 
of the O5e star HD 46223, which has a line-of-sight 
extinction A$_V \sim 2.1$. Luminosity effects are explored with 
spectra of the B7III star HD 35497 and the B7V star HD 43153. 
The H$\alpha$ line profile in the mean blue spectrum is more akin 
to that in B giants as opposed to B dwarfs.}
\end{figure}

	There is also significant interstellar Na D absorption in the spectra of 
stars in the bright sample. The spectrum of the O5e star HD46223 from the Le 
Borgne et al. (2003) compilation is shown in Figure 13. This star has a 
line-of-sight extinction A$_V \sim 2.1$, and 
Na D absorption in its spectrum has a depth that is not greatly 
different from that of the stars in the bright sample. 
There are only modest star-to-star differences in the depth of Na D 
in Figure 11, and this likely reflects the similar colors (and hence reddenings) of the 
stars in this sample. There is greater dispersion in the depth of 
Na D among the stars in the red sample, with Star 14 showing the deepest Na D 
feature among the bright early-type stars.

	Star-to-star variations in the depth of the 6494\AA\ 
DIB are evident in the right hand panel of Figure 12. This 
DIB is weakest in the two stars with H$\alpha$ emission, but can be seen in 
the spectra of Stars 05 and 14. The depth of the 6494 DIB tends to correlate only 
weakly with those of other DIBs (Cami et al. 1997). Cami et al. (1997) also 
find a possible anti-correlation with the 6613\AA\ feature, which is marked 
in the right hand panel of Figure 12.

	Equivalent width (EW) measurements of Na D were 
made from mean spectra that were constructed for three different groupings: 
(1) all of the blue sub-sample stars, (2) the 
mean spectrum of Stars 4, 5, and 14, which are the stars in the red sub-sample that 
do not show H$\alpha$ emission and are not foreground dwarfs, and (3) the mean 
spectrum of all stars in the bright sample, but again excluding the H$\alpha$ emission 
Stars 29 and 31 and the foreground dwarf Star 32. The results 
are listed in Table 4. The uncertainties in Table 4 reflect the dispersion in EWs 
found by perturbing the start and end points of the EW measurements.

\begin{deluxetable}{lc}
\tablecaption{Na D equivalent widths (EWs) 
\\ in mean spectra constructed \\ from the Bright sample.}
\startdata
\tablewidth{0pt}
\tableline\tableline
ID & EW \\
\# & (\AA\ ) \\
\tableline
Blue Sub-sample & $1.5 \pm 0.1$ \\
Stars 4, 5, \& 14 & $1.9 \pm 0.1$ \\
Bright Sample & $1.6 \pm 0.1$ \\
(No Stars 29, 31, \& 32) & \\
\tableline
\enddata
\end{deluxetable}

	The depths of the Na D lines in the spectra of early type stars 
can be used to estimate their reddening. Poznanski et 
al. (2012) find a relation between the EW of the Na D lines and E(B--V). 
A blind application of equation 9 of Poznanski et al. (2012) to 
the Na D EW measured for the early-type stars in the blue sub-sample 
yields E(B--V) $= 0.80 \pm 0.25$. Applying the Na D EW from the entire 
bright sample using the entry in the last row of Table 4 yields E(B--V) = $1.05 
\pm 0.25$. These values agree with the reddening measured from the CMD. There is thus 
consistency between the $g'-i'$ color of the MS and the depth of the Na D feature 
in the spectra of early-type stars in the NGC 3105 field.

\subsection{Intermediate Magnitude Stars}

	Spectra centered on Na D and H$\alpha$ of stars in 
the intermediate brightness sample are compared in Figure 14. 
The spectra have been smoothed to a resolution of 9.5\AA\ to 
boost the S/N ratio while maintaining a spectral resolution that is 
sufficient to allow strong absorption features to be identified (e.g. Worthey 
et al. 1994; Pickles 1998). Star 11, which is displayed at the 
bottom of the figure, was observed because it could be placed 
in the same slit as a higher priority target.

\begin{figure}
\figurenum{14}
\epsscale{1.2}
\plotone{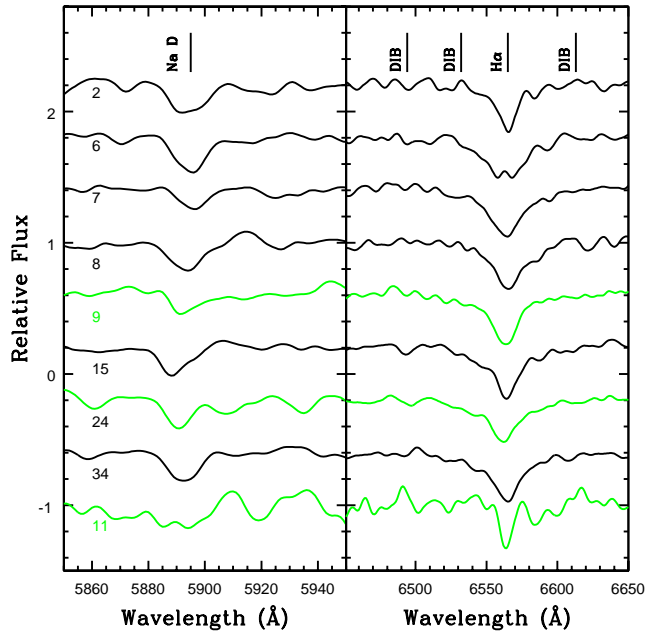}
\caption{Spectra near the Na D and H$\alpha$ lines of intermediate brightness stars. 
The spectra shown in black are of stars in Fields 1 and 2, while the green spectra are 
of stars in the Shoulder region. The spectra have 
been smoothed to a 9.5\AA\ resolution. The spectrum of Star 11, 
which has a red color and was observed because it could be placed in the same slit as 
a higher priority target, is shown at the bottom of the figure. 
The majority of these stars show broad H$\alpha$ absorption that is characteristic of 
A stars. The locations of the DIBs marked in Figure 11 are shown.}
\end{figure}

	There is a range of H$\alpha$ characteristics in Figure 14. 
Star 6 has H$\alpha$ in emission, while Star 11 has a narrow H$\alpha$ 
profile. The other stars have broad H$\alpha$ profiles that 
are reminiscent of A spectral-types. Despite smoothing, 
considerable noise remains in the spectra near Na D, with 
Star 11 having the lowest S/N ratio near 5900\AA\ 
because of its very red color. Star 6 has the deepest Na D absorption. 
Circumstellar material may contribute to Na D absorption in the spectrum of this star
given the presence of H$\alpha$ emission in its spectrum. 

	The spectra of stars 9 and 24, which are in the Shoulder region, were 
combined, as were the spectra of Stars 2, 7, 8, 15, and 34 -- Star 6 was not 
considered when computing the Shoulder region mean given the presence of H$\alpha$ 
emission. Stars from the Le Borgne et al. (2003) library are compared with the 
combined GMOS spectra in Figure 15. The reference spectra have been smoothed to 
the same resolution as the GMOS data. The wavelength interval shown in Figure 
15 contains well-calibrated probes of metallicity and surface gravity 
(e.g. Worthey et al. 1994). 

\begin{figure}
\figurenum{15}
\epsscale{1.2}
\plotone{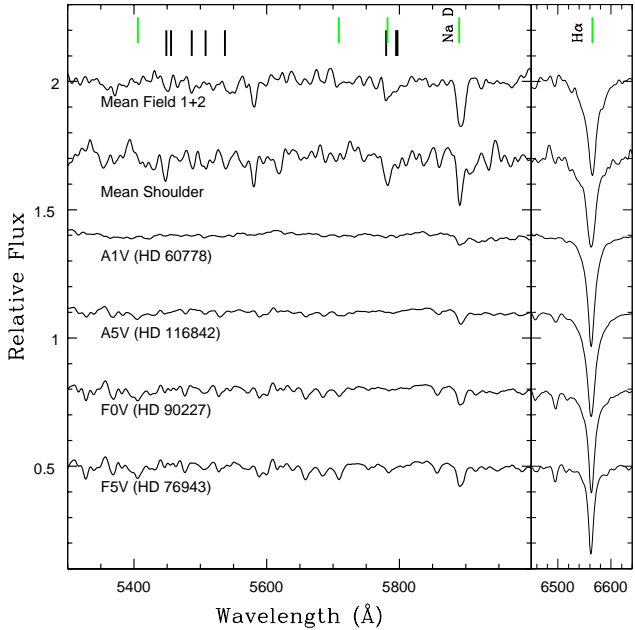}
\caption{Mean spectra of intermediate magnitude stars 
in Field 1$+$2 and the Shoulder regions. The spectra have been 
smoothed to a spectral resolution of 9.5\AA. Spectra of stars from the Le Borgne 
et al. (2003) library that have been normalized to the continuum and smoothed 
to match the resolution of the GMOS spectra are also shown. 
DIBs that have mean EWs $> 0.1$\AA\ in the Diffuse Interstellar Band Catalogue 
(http://leonid.arc.nasa.gov/DIBcatalog.html) are indicated, and 
the green lines mark the locations of Fe absorption features at 5335, 5406, 5709, 
and 5782\AA. Efforts to estimate spectral-types at wavelengths $< 6000\AA$ 
in these spectra are complicated by noise and interstellar absorption features. 
However, the depth and shape of H$\alpha$ absorption in the Shoulder 
and Field spectra is consistent with a mid-A spectral type.}
\end{figure} 

	The Na D line in both mean spectra is deeper than in the bright star 
sample, suggesting that there is a photospheric contribution to 
Na D absorption in the intermediate magnitude sample.
Efforts to classify the mean GMOS spectra are complicated 
by noise and interstellar absorption features at wavelengths $< 6000$\AA, 
and so we rely on H$\alpha$ to estimate spectral-types. 
The depth and shape of H$\alpha$ in the median cluster and 
field spectra are consistent with a spectral type $\sim$ A5V. This was confirmed by 
subtracting reference spectra from the mean GMOS spectra and examining the residuals. 

	A distance modulus can be computed if it is assumed that the stars in the 
intermediate brightness sample follow the solar-neighborhood M$_{V}$ $vs.$ 
spectral-type relation. Jaschek \& Gomez (1998) examine the M$_V$ calibration of 
spectroscopic standards, and their Table 1 suggest that M$_V = 1.7 \pm 0.3$ 
for an A5V star. The uncertainty reflects the difference in M$_V$ for the two A5V 
stars in the Jaschek \& Gomez sample. The stars in the intermediate magnitude 
sample have $i' \sim 18.7$, and if E(B--V) = 1 this then corresponds to $V 
\sim 19.3$, or $V_0 \sim 16.2$. The distance modulus of the star in the 
intermediate sample is then $\sim 14.5 \pm 0.3$. This agrees with the values found from 
the CMDs in Section 4. While it is not clear if
the stars in the intermediate sample belong to NGC 3105, 
they appear to be at a similar distance.

\subsection{Faint Stars}

	If the stars in the faint sample are on the 
MS of NGC 3105 then the isochrones suggest that they should have a 
G spectral-type. It is difficult to estimate the spectral-types of 
individual stars in the faint sample as the spectra 
have a low S/N ratio. However, Na D is a deep absorption feature in the 
spectrum of late-type dwarfs, and it is seen in many of the faint sample spectra. 
Na D provides a means of identifying foreground dwarfs, 
as these objects will have lower masses than stars in NGC 3105, and so 
their spectra will have deeper Na D absorption.

	Even though the majority of objects in the faint 
sample were intentionally selected to be in the 
Cluster or Shoulder regions to boost the probability of cluster membership, 
source counts suggest that roughly half of the stars in the 
faint sample will not be cluster members (Figure 6). 
In fact, four stars in the faint sample (\#'s 10, 19, 25, and 26) were found to have 
Na D absorption that was much deeper than that in the other objects, and these were 
flagged as candidate field stars. A median spectrum of the stars in the faint sample, 
but excluding \#'s 10, 19, 25, and 26, was constructed. The median was taken 
in an effort to suppress the influence of any field stars that might remain in 
the sample. The resulting spectrum, smoothed to a resolution of 400 to boost 
the S/N ratio, is shown in Figure 16. Spectra of G and K stars from the Le Borgne 
et al. (2003) compilation, smoothed to the same resolution, are also shown 
in that figure.

\begin{figure}
\figurenum{16}
\epsscale{1.2}
\plotone{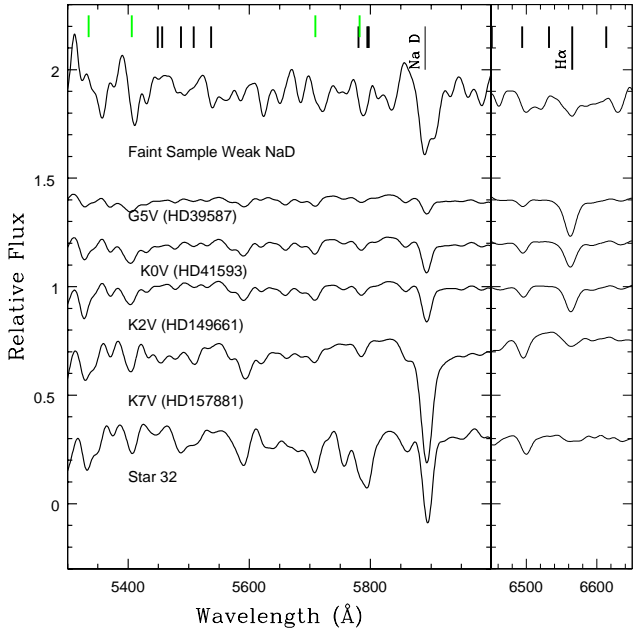}
\caption{Median spectrum of stars in the faint sample -- stars 
that have strong Na D absorption, and thus are suspected foreground dwarfs, have 
been excluded. Also shown are spectra of G and K stars from 
the Le Borgne et al. compilation, and the smoothed spectrum of Star 32. 
All spectra have a resolution $\lambda/\Delta \lambda = 400$. 
The locations of strong DIBs and Fe absorption lines from Figure 15 are also indicated. 
The depths of Na D and H$\alpha$ suggests a K spectral type for the majority 
of stars that make up the faint star sample. Star 32 from the bright sample 
likely has a spectral type in the range K5V -- K7V, and its spectrum 
has features in common with some of those in the median faint sample spectrum.}
\end{figure} 

	H$\alpha$ emission might be expected if there was large-scale chromospheric 
activity among the stars in the faint group, and this is not seen 
in the median spectrum. Rather, there appears to be weak H$\alpha$ 
absorption in the combined spectrum. As for Na D, this feature has an EW of $\sim 
7.8$\AA\ in the median faint sample spectrum. After accounting for interstellar 
absorption, Na D in the median spectrum is much deeper than expected in G 
dwarfs, and is consistent with an early to mid-K spectral-type. 

	There are interstellar absorption features in the median spectrum
that complicate spectral classification based on comparisons with stars that 
are subject to only modest levels of extinction. As a late-type 
foreground dwarf that is also heavily extincted, Star 32 is thus an 
interesting comparison object for the faint sample median spectrum. 
The spectrum of Star 32, smoothed to a resolution of 400, is 
shown in Figure 16. There are many similarities between the median faint sample 
spectrum and that of Star 32. The lack of H$\alpha$ absorption in the Star 32 spectrum 
suggests that it has a later spectral-type than the majority of the faint sample 
members. Based on the comparisons in Figure 16 it appears that Star 
32 has a spectral-type in the range K5V -- K7V. 

	The deep Na D and weak H$\alpha$ absorption in the median faint sample spectrum 
is perhaps surprising given that a G spectral type was expected for these stars. 
Star spot activity has the potential to affect the effective temperature of 
stars if the spots fall over a large part of the stellar disk. 
Spotted single stars will have lower effective temperatures than unspotted 
models, and models examined by Somers \& Pinsonneault (2015) suggest that 
the impact on effective temperature is $\sim 200 - 300$ K. This is not sufficient to 
produce the difference between the observed and expected spectral types. 
Large-scale spot activity might also produce chromospheric emission, and no evidence 
for this is seen. A more pragmatic explanation is that -- the culling of obvious 
field stars based on Na D depth notwithstanding -- foreground stars dominate 
the sample. Spectra of stars with colors and magnitudes 
like those in the faint sample with a higher S/N ratio will be 
useful to distinguish between cluster and field stars at this brightness in NGC 3105.

\section{SUMMARY \& DISCUSSION}

	Images and spectra obtained with GMOS on GS have been used to investigate 
the age, distance, and reddening of the outer Galactic disk open cluster NGC 3105. 
The $g'$ and $i'$ GMOS images allow photometry to be obtained of sources 
that are much fainter than in previous studies at visible/red wavelengths, reaching 
magnitudes where cluster PMS stars might be expected if NGC 3105 has an 
age $\sim 20$ Myr, as suggested by some age estimates. 
The spectra enable the first detailed reconnaisance of the stellar content 
of NGC 3105 and the surrounding field below the MSTO, and yield independent 
reddening and distance estimates. 

	There is good agreement between the reddening and distance 
found from the photometry and spectra. The reddening estimated from the locus 
of MS stars on the CMD agrees with that found from the depth of Na D absorption 
in the spectra of bright early-type stars in and around NGC 3105 using a 
calibration found by Poznanski et al. (2012). The distance estimated 
from MS-fitting is consistent with that found in previous studies, and also agrees 
with the distance estimated from the spectral-types of candidate MS stars at 
intermediate magnitudes. We note that these stars are located throughout the 
GMOS science field, and so are not restricted to the dense inner regions of 
the cluster where membership probabilities are high.

	An age of at least 32 Myr is estimated for NGC 3105 
based on the photometric properties of (1) the brightest 
stars near the cluster center, which include MSTO and post-MS objects, and (2) the 
morphology of the lower parts of the CMD. An upper limit to the age range is 
71 Myr, and this is defined by the photometric properties of candidate upper MS stars. 
The isochrones suggest that the GMOS observations do not sample PMS stars 
in NGC 3105, and so deeper observations will be required to find these 
objects at visible/red wavelengths. This may prove to be challenging as 
it can be difficult to identify stars with sub-solar masses in low latitude clusters
as foreground/background stars may have similar photometric properties. 
In the case of very young clusters individual PMS stars that are surrounded 
by accretion discs may be identified based on their distinct photometric 
and spectroscopic properties. However, the circumstellar 
discs that produce these signatures have a characteristic decay time of 
a few Myr (Fedele et al. 2010, but see also e.g. De Marchi et al. 2011a), 
negating their usefulness as beacons for identifying sources in clusters 
as old as NGC 3105. Still, Scicluna et al. (2014) suggest that in some cases 
circumstellar discs may be rejuvenated by the accretion of fresh molecular material, 
potentially increasing the time interval over which disks may be detected.
In any event, ensembles of candidate stars can be identified statistically by examining 
the spatial distribution and photometric properties of objects (e.g. Davidge 2015), 
as is done here.

	The consistency of age estimates based on observations 
that span a range of wavelengths are an important test of stellar structure models. 
Davidge (2014) used near-infrared observations 
to examine the CMD and LF of NGC 3105 in the sub-solar mass regime. 
The $(K, J-K)$ CMD of the central $85 \times 85$ arcsec of NGC 3105 
constructed by Davidge (2014) shows a well-defined sequence at faint 
magnitudes that was attributed to the PMS. The PMS sequence in the $(K, J-K)$ CMD is 
bracketed by 20 and 40 Myr isochrones from Bressan et al. (2012). 
The age of NGC 3105 estimated from the PMS locus in that CMD 
is 25 -- 30 Myr. {\it Considering only the faintest stars in the 
GMOS CMD}, the comparisons in Figure 6 suggest that NGC 3105 has an age $> 22$ Myr. 
The ages derived from PMS/faint MS stars in the near-IR and at visible/red 
wavelengths are thus consistent.

	Davidge (2014) also examined the $K$ LF of the center of NGC 3105 to a faint 
limit $K \sim 19.5$, which corresponds to $\sim 0.5$M$_{\odot}$. The flat shape of 
the $K$ LF at the faint end is similar to the $i'$ LF, which is flat from $i' \sim 18$ 
($\sim 3.5$M$_{\odot}$) to $i' \sim 22$ ($\sim 0.8$M$_{\odot}$), 
and is matched by models that have a Chabrier (2003) MF. 
However, Davidge (2014) found that model LFs that are based on the Bressan 
et al. (2012) isochrones and that assume a solar neighborhood MF do not reproduce 
the flat nature of the $K$ LF at faint magnitudes. The $K$ observations probe 
less massive stars than the GMOS $i'$ data. If the 
interpretation of the $K$ LF is correct then a downturn in the $i'$ LF will occur 
1 -- 2 magnitudes below the faint limit of the GMOS data.

	There is excess MIR emission associated with some of the brightest stars in 
the GMOS science field, as expected if these objects have circumstellar disks. 
The spectra of four of the stars in the bright and intermediate 
brightness samples have H$\alpha$ in emission, which is 
not surprising as line emission is common among early-type stars 
in systems with ages $> 10$ Myr. If these stars are not 
cluster members then they may have formed more-or-less concurrently with 
NGC 3105 in structures that have dispersed into the field. 
Simulations of star formation in molecular clouds discussed by
Bonnell et al. (2011) suggest that both clustered and diffuse components
form. While many young star-forming regions will expand and dissipate, some retain 
primordial sub-structures (e.g. Gregorio-Hetem et al. 2015), and NGC 3105 
may be such a remnant. Diffusely distributed young populations might be expected to 
survive longer in the low density regions of the outer Galactic disk than in 
higher density areas at smaller Galactocentric radii.

	The presence of a diffuse bright population notwithstanding, 
star counts indicate that the brightest, most massive members of NGC 
3105 are more centrally concentrated than the fainter, low mass members, indicating 
that there is mass segregation. The central concentration of bright members 
might have a primordial origin. Simulations (e.g. Dale et al. 2015) suggest 
that -- in the absence of feedback -- star formation 
rates in forming clusters tend to be highest in regions where the potential well is 
deepest. In fact, evidence for primordial segregation is seen in the Orion star-forming 
region (e.g. Hillenbrand \& Hartmann 1998), and simulations suggest that 
it can also occur in very low mass clusters (Kirk et al. 2014). 
While feedback can suppress star formation in dense regions
(e.g. Dale et al. 2015), clusters that have a low stellar mass will have 
lower probablities of forming the massive stars that drive feedback than more
massive systems, and so long-lived central knots may form preferentially in 
low mass clusters. There is an observational bias favoring the detection of 
low mass clusters that have bright, central concentrations, and so the observed 
frequency of such systems may overestimate their actual numbers.

\clearpage

\end{document}